\documentstyle[manuscript,aps,psfig,12pt]{revtex}
\input epsf
\tightenlines
\begin{document}
\preprint{SWAT/97/162}
\draft
\title{Magnetic Monopoles as Agents of Chiral Symmetry 
Breaking in U(1) Lattice Gauge Theory}
\author{Tom Bielefeld}
\address{Institut f\"ur Theoretische Physik, Universit\"at Bremen,\\
Postfach 330 440, D-28 334 Bremen, Germany}
\author{Simon Hands}
\address{Department of Physics, University of Wales Swansea,\\
Singleton Park, Swansea, SA2 8PP, U.K.}
\author{John D. Stack}
\address{Department of Physics,
University of Illinois at Urbana-Champaign\\
1110 West Green Street,
Urbana, IL 61801-3080, U.S.A.}
\author{Roy J. Wensley}
\address{Department of Physics and Astronomy, Saint Mary's College,\\
Moraga, CA 94575, U.S.A.}
\maketitle

\begin{abstract}
We present results suggesting that magnetic monopoles can account
for chiral symmetry breaking in abelian gauge theory. 
Full $U(1)$
configurations from a lattice simulation are factorized into magnetic
monopole and photon contributions. The expectation
$\langle\bar{\psi}\psi\rangle$ is computed using the monopole configurations
and compared to results for the full $U(1)$ configurations. It is shown that
excellent agreement between the two values of $\langle\bar{\psi}\psi\rangle$
is obtained if the effect of photons, which
``dress" the composite operator $\bar\psi\psi$, is included.
This can be estimated
independently by measurements of the physical fermion mass in the photon
background.
\end{abstract}
\vfill
\pagebreak

$U(1)$ lattice gauge theory provides an ideal laboratory for studying the
effects of monopoles.  In $U(1)$, all non-perturbative effects are known to 
be caused
by monopoles \cite{Kogut}.  
Further, there is an efficient procedure for identifying 
the magnetic current of monopoles in the link angle configurations 
generated in a $U(1)$ lattice gauge theory simulation \cite{DeGT}.  Previously,
two of the present authors have studied the heavy quark potential in
$U(1)$ lattice gauge theory, and shown that in the confined phase,
the value of the string tension
calculated from monopoles agrees quantitatively with the full $U(1)$
string tension calculated directly from link angles \cite{Stack}.

In this paper, we turn our attention to the quenched chiral condensate
in $U(1)$ lattice gauge theory, $\langle\bar\psi\psi\rangle$.
We expect (and find) that this has a non-zero value in the confined phase,
and vanishes in the deconfined phase.  Our interest is in how the value of
$\langle\bar\psi\psi\rangle$ calculated from the link angle configurations
compares with the same quantity calculated from monopoles.  Ascribing
non-perturbative effects to monopoles would say that in the confined phase,
a non-vanishing condensate
will be induced by monopoles.  Nevertheless, since a matrix element of
a product of field operators is involved, this value should be renormalized by
short distance perturbative effects {\it i.e.}
photons, so to make a quantitative comparison, we also 
need to calculate this renormalization.  Our method for doing
so, described in more detail below, is to compute the ratio of renormalized to
bare masses for a charge moving solely in the field of photons.  Our principle
result is that applying  
this renormalization factor to the monopole
value of $\langle\bar\psi\psi\rangle$ yields the full $U(1)$ answer, to within
statistical errors.

Let us now briefly describe the factorization of the link variables into
photon and monopole parts.  We start by resolving the $U(1)$ plaquette
angles $\phi_{\mu\nu}$ into fluctuating and monopole parts,
$$\phi_{\mu\nu}=\phi'_{\mu\nu}+2\pi m_{\mu\nu},$$  where
$\phi'_{\mu\nu} \in (-\pi,+\pi]$. From the monopole
term $m_{\mu\nu}$,
we can define two currents, one electric, the other magnetic.  The
electric current is defined by $j_{\mu}=\nabla^{-}_{\mu}m_{\mu\nu}$.
The magnetic current of monopoles 
is defined by $m_{\mu}=\nabla^{+*}_{\mu}
m_{\mu\nu}$, where $^{*}m_{\mu\nu}$ is the dual of $m_{\mu\nu}$.  The
currents $j_{\mu}(m_{\mu})$ reside on direct(dual) lattices.

We will compute the chiral condensate using the Banks-Casher formula
\cite{Banks}
(see below) which involves the density of eigenvalues of the Dirac
operator.  Since the Dirac particle is electrically charged, the effect
of monopoles on it must be represented by an electric vector potential.
This four-vector potential derives from the electric current $j_{\mu}$, 
and is analogous to the
familiar three-vector potential $\vec{A}$ used for the case of
a static monopole in the continuum.  We have
\begin{equation}
A_\mu^{mon}(x)\equiv g\sum_y v(x-y)j_\mu(y),
\end{equation}
where $g=2\pi/e$ is the magnetic unit of charge, 
and $v$ the lattice Coulomb propagator in Feynman gauge,
satisfying $\nabla^+_\mu\nabla^-_\mu
v(x-y)=-\delta_{x,y}$. By defining
$U_\mu^{mon}(x)\equiv \exp(ieA_\mu^{mon}(x))$,  we define the
photon link by demanding that 
$\{U_\mu\}$ be factorized on each link:
\begin{equation}
U_\mu(x)\equiv U_\mu^{mon}(x)U_\mu^{phot}(x).
\end{equation}

We can now examine the chiral condensate $\langle\bar\psi\psi\rangle$
separately on configurations $\{U_\mu\}^{mon}$ and
$\{U_\mu\}^{phot}$, as well as the full $\{U_\mu\}$. 
In the quenched approximation 
$\langle\bar\psi\psi\rangle$ can be defined in the chiral limit
via the eigenvalues of the
Dirac operator~\cite{Banks}:
\begin{equation}
\langle\bar\psi\psi\rangle={\pi\over V}\rho(\lambda=0),
\end{equation}
where $V$ is the lattice volume and $\rho(\lambda)d\lambda$ is the
number of eigenvalues satisfying $D{\!\!\!\! /}\,\vert n\rangle=
i\lambda_n\vert n\rangle$ in the interval $(\lambda,\lambda+d\lambda)$,
with $D{\!\!\!\! /}\,[U_\mu]$ the staggered lattice fermion
kinetic operator
\begin{equation}
D{\!\!\!\! /}\,_{xy}[U_\mu]={1\over2}\sum_\mu
(-1)^{x_1+\cdots+x_{\mu-1}}
\left(U_\mu(x)\delta_{y,x+\hat\mu}-U_\mu^*(y)\delta_{y,x-\hat\mu}\right).
\end{equation}
For finite $V$, $\rho(0)$ vanishes; therefore an extrapolation to the
$\lambda\to0$ limit must be made from the spectrum density at non-zero
$\lambda$. In practice $\rho(\lambda)$ can be estimated by measuring the 
lowest $O(25)$ eigenvalues of $D{\!\!\!\! /}\,$ per configuration
using the Lanczos
algorithm, and then using a binning procedure~\cite{Lanczos}.

Calculations of the chiral condensate were performed with $U(1)$
configurations generated using the standard Wilson action on a $12^4$
lattice.  Three different values of the inverse coupling $\beta$ were
chosen: $\beta=1.005, 1.010,$ and $1.020$. For each value of
$\beta$, 5000 lattice updates were performed before beginning measurements 
to allow for equilibration.  The next 4000 configurations were used for
measurements skipping every 20 lattice updates to reduce correlations.
For values $\beta \le 1.010$,
our results exhibited the breaking of chiral symmetry (i.~e.
$\langle\bar\psi\psi\rangle\ne0$).  These two values of
$\beta$ are known to correspond to the confinement phase as
well\cite{gupta}.
For $\beta=1.020$ the configurations were found to be in the chirally
symmetric
phase, which is coincident with the Coulomb phase.

Our results for the chiral condensate calculations for the
chirally broken phase are presented in
Figs.~(1) and (2). We show both the full $U(1)$ calculations and the
results from the monopole gauge field configurations obtained using
Eq.~(1). Using a linear fit and Eq.~(3), the value of the chiral
condensate was extracted
from the 10 lowest values of $\rho(\lambda)$.
The values for $\langle\bar\psi \psi\rangle$ are shown in Table~I.

In Fig.~(3) the results for the chiral condensate from full $U(1)$
fields
and monopoles are shown for  $\beta=1.020$.  For comparison, the
results from the $\beta=1.010$ monopole configurations are also
included.
Linear fits to $\rho(\lambda)$ yield a very small, but finite
intercept.  However, the values are found to be about a factor of 100
times smaller than the results in the broken phase.  Thus, we are
confident
of being in the chirally symmetric phase for $\beta=1.020$.  For
completeness, the $\beta=1.020$ results are also included in Table~I.

In Fig.~(4) we show the eigenvalue spectrum calculated using the
$\{U_\mu\}^{phot}$ background from both broken and symmetric phases.
There is no signal of chiral symmetry breaking
(note that we choose anti-periodic
boundary conditions for the fermions in the temporal direction to
avoid zero modes from near-plane wave solutions).
This supports our conjecture that chiral symmetry
breaking in this model can be ascribed entirely to monopoles.

Is it possible to account for the mismatch between 
$\langle\bar\psi\psi\rangle^{U(1)}$ and
$\langle\bar\psi\psi\rangle^{mon}$? Since $\bar\psi\psi$ is a
composite field operator, we expect it to be modified by quantum corrections
independently of whether it acquires a vacuum expectation value. In
perturbation theory, UV fluctuations in general result in the requirement for
all field operators to be renormalized, and for
composite operators to have an additional renormalization.
This consideration leads us to the following hypothesis: the mismatch
between the two condensate measurements is due to the rescaling of
the $\bar\psi\psi$ operator by the fluctuations of the gauge fields
contained in the $\{U_\mu\}^{phot}$ configurations; more concisely
\begin{eqnarray}
\langle\bar\psi\psi\rangle^{mon} &=&
\langle(\bar\psi\psi)^{tree}\rangle^{mon} \nonumber\\
\langle\bar\psi\psi\rangle^{U(1)} &=&
\langle(\bar\psi\psi)^{phot}\rangle^{mon}, 
\end{eqnarray}
where $(\bar\psi\psi)^{phot}$ is the dressed operator which takes
into account photon-like fluctuations, and the expectation value is
taken in the monopole-only backgrounds. If the hypothesis is correct,
then we can write
\begin{equation}
(\bar\psi\psi)^{phot}=Z(\bar\psi\psi)^{tree},
\end{equation}
ie. the tree-level operator is multiplicatively enhanced by the photon
fluctuations, the condensate data suggesting that the factor
$Z\simeq1.5$.

In support, we now describe an alternative and independent determination
of $Z$, obtained by measuring the physical fermion mass $m_R$
in the $\{U_\mu\}^{phot}$ background. On the assumption that the
fluctuations in $\{U_\mu\}^{phot}$ are approximately Gaussian, then
the dressing of the $\bar\psi\psi$ operator is given by a set of
Feynman diagrams. Let $\Sigma(p)$ denote the complete set of 1PI
diagrams describing corrections to the fermion two-point function. Then
for the dressed fermion propagator we have
\begin{equation}
S_F^{phot}(p)={1\over{ip{\!\!\! /}\,+m_0-\Sigma(p)}}
\simeq{{Z_2(a,\mu)}\over{ip{\!\!\! /}\,+m_R}},
\label{eq:dressedprop}
\end{equation}
where $m_0$, $m_R$ denote bare and physical fermion masses
respectively, $Z_2$ is a wavefunction rescaling which by analogy with
perturbative QED we expect to be both gauge and cutoff-dependent,
and the $\simeq$ symbol shows that the second equality holds only in 
the neighborhood of some subtraction point $ip{\!\!\! /}\,=\mu$.
Equation (\ref{eq:dressedprop}) may be rearranged to read
\begin{equation}
\Sigma(\mu)=(1-Z_2^{-1}Z_m)m_0+\mu(1-Z_2^{-1})
\label{eq:Zmdef}
\end{equation}
with 
\begin{equation}
m_R=Z_m(a,\mu)m_0.
\end{equation}
Now, using the identity
\begin{equation}
{d\over{dm_0}}{1\over{ip{\!\!\! /}\,+m_0}}={{-1}\over
{(ip{\!\!\! /}\,+m_0)(ip{\!\!\! /}\,+m_0)}},
\end{equation}
we see that the operation of differentiating with respect to $m_0$
is equivalent to a zero momentum insertion of a $\bar\psi\psi$ operator
in a Feynman diagram. Therefore $-d\Sigma/dm_0$ is the set of 1PI
diagrams, having one external $\psi$ and one external $\bar\psi$, 
which describe corrections to $\bar\psi\psi$, ie. 
\begin{equation}
(\bar\psi\psi)^{phot}=Z_2\left(1-{{d\Sigma(\mu)}\over {dm_0}}
\right)
(\bar\psi\psi)^{tree}\equiv Z(\bar\psi\psi)^{tree},
\end{equation}
ie.
\begin{equation}
ZZ_2^{-1}=1-{{d\Sigma}\over{dm_0}},
\label{eq:Zdef}
\end{equation}
where $Z(a,\mu)$ is the rescaling factor we seek, and the factor
$Z_2$ is included because in the diagrammatic approach
matrix elements are 
evaluated with dressed fermion propagators on the external legs.
Combining (\ref{eq:Zmdef}) with (\ref{eq:Zdef}), we find
\begin{equation}
Z=Z_m={m_R\over m_0}.
\end{equation}
Hence a measurement of $m_R$ in the $\{U_\mu\}^{phot}$
background as a function of $m_0$ gives
an independent estimate of $Z$.

This argument is essentially the same as that used to establish
$Z_{\bar\psi\psi}=m_0/m_R$ in standard renormalized perturbation 
theory, where $(\bar\psi\psi)_R=Z_{\bar\psi\psi}(\bar\psi\psi)$ is
the renormalized operator whose Green functions with other renormalized
fields and operators are finite~\cite{Collins}.
Here we make no attempt to define a renormalized operator, since
bare quantities are evaluated in a lattice simulation, but rather use
the same formalism to quantify the effects of operator enhancement
by quantum fluctuations. Note that we have not specified the
subtraction point defining $Z$ very precisely, and have
assumed that $Z$ is independent of $m_0$. From experience in
perturbation theory, we expect that numerically the most significant
contribution to $Z$, of $O(\ln a)$,
comes from short wavelength fluctuations, and that $Z$ is relatively
insensitive to the details of the subtraction.
 
To measure the physical fermion mass in the $\{U_\mu\}^{phot}$
background, we performed calculations of the fermion propagator
starting with the same configurations used to compute $\rho(\lambda)$. 
Since the
fermion propagator is not gauge invariant, it is first necessary to fix
a gauge. 
Although a gauge transformation of $\{U_\mu\}^{phot}$
is not strictly a symmetry of the full theory, this procedure is
justified since $\bar\psi\psi$, and by hypothesis $m_R$, are gauge
invariant.
In this work we have used Landau gauge \cite{codd}
and extracted the lattice fermion timeslice
propagator 
\begin{equation}
C_f(x_4)=\mbox{Re}\sum_{x_1,x_2,x_3 even} \biggl\langle (D{\!\!\!\!
/}\,[U_\mu]+m_0)_{0,x}^{-1}\biggr\rangle^{phot}
\end{equation}
using a conjugate gradient routine, for bare mass values
$m_0a=0.1,0.09,\ldots,0.04$. The restriction to spatial sites
an even number of lattice spacings from the origin in each direction
improves the signal~\cite{Gock}. Note that we have not attempted to fix
the residual gauge freedom as in \cite{Gock}, but have instead
relied on the fluctuations in $\{U_\mu\}^{phot}$ being intrinsically
small (see below). 
The physical mass $m_R$ can now be estimated by fitting $C_f$
to the following functional form: 
\begin{equation}
C_f(x_4)=A\left(\exp(-m_Rx_4)+(-1)^{x_4}\exp(-m_R(L-x_4))\right),
\end{equation}
where $L$ is the lattice size in the time direction. We used
100 configurations from each of the three $\beta$ values previously
studied.
In order to
take into account correlations between renormalized mass estimates due
to using the same configurations for different bare masses we applied a
bootstrap fitting routine.

To determine $Z$ we plotted the mass obtained from $C_f$
against the input bare mass, and fitted the results to a linear
form:
\begin{equation} \label{eq:xxx}
m_R (m_0) = Z \; m_0 + b.
\end{equation}
Fig.~(5) shows a graph of $m_R(m_0)$ from our simulation for
$\beta=1.010$.
The graphs for the other two values of $\beta$ are very similar.  The
results for $Z$ and $b$ from the linear fits for each value of $\beta$
are shown in Table~II.  From the table
it can be observed that the value of $Z$ is a  smooth function through
the phase transition.  Unexpected in our results were the small
non-zero values of the intercept.  This will be discussed briefly
below.

Finally, we compare our value of
$Z$ calculated using $\{U_\mu\}^{phot}$  with the value observed
in the chiral condensate calculations.  We define the
renormalization factor from the chiral condensate to be
\begin{equation}
\label{eq:psi}
Z_{\bar{\psi}\psi}={\langle\bar{\psi}\psi\rangle^{ U(1)} \over
\langle\bar{\psi}\psi\rangle^{mon}},
\end{equation}
and present the comparison in Table~III.  The two independent
determinations
of the renormalization factor show excellent agreement.  This answers
the question asked earlier: Is it possible to account for the mismatch
between $\langle\bar\psi\psi\rangle^{U(1)}$ and
$\langle\bar\psi\psi\rangle^{mon}$?  Our results demonstrate
that the mismatch is simply
operator renormalization due to the photons.  

As yet we have no satisfactory explanation for the non-zero intercepts
in Table~II, implying a small breaking of chiral symmetry in the
photon-only configurations. It could be either a finite volume effect,
or perhaps a residual gauge freedom associated with 
Gribov copies and the consequent non-uniqueness of the Landau gauge, or
perhaps due to spatially constant gauge field modes. 
The only effect that we have investigated
quantitatively is that of constant modes.  We repeated our previous 
calculations at $\beta=1.005$, this time performing a global gauge 
transformation
to remove constant modes from the gauge fields.  
This was done in the following way:
The full gauge fields were shifted by a direction dependent constant,
\begin{equation}
A^{\prime}_\mu(x) = A_\mu(x)+c_\mu,
\end{equation}
where $c_\mu$ is chosen so $\sum_x A^{\prime}_\mu(x)=0$.  The new
gauge fields $A^{\prime}_\mu$ were used to recompute $Z$ and the
chiral condensates.
In repeating the calculations
with this gauge fixing we observed no measurable difference in the 
results.

Although we are not able to explain the small non-zero intercepts, 
the overall agreement of the slope with the value of $Z$ required to
explain the condensate results is extremely satisfying,
and supports our hypothesis on the role of the residual photon
fluctuations enhancing the $\bar\psi\psi$ operator.

 An obvious direction in which to extend this analysis is
the exploration of chiral symmetry breaking in non-abelian gauge
theories following the identification of monopole networks after 
abelian projection. Studies of this kind have already appeared~\cite{LeeWen};
it is interesting to note that in the maximal abelian gauge used, 
there is a similar mismatch in the measured values
of $\langle\bar\psi\psi\rangle$ between full and monopole-only
configurations. It would be interesting to check whether
this effect could be accounted
for by the operator renormalization described here.

Finally note that even though this argument assumed a split of
the background gauge field into
Gaussian fluctuations, described by Feynman diagrams, and
non-perturbative monopole-only configurations, the idea of classifying
effects into those which rescale a local operator and those which lead
to a non-vanishing expectation value for the operator may be
generalized. For instance, it may in principle be possible to extend the 
analysis by factoring configurations in a scale dependent fashion,
including only large monopole loops in the monopole part, and 
seeing if small monopole loops simply renormalize $\bar\psi\psi$
by comparing the $Z$ factors measured in two different ways.
This may prove to be an effective probe of
the scale at which non-perturbative effects manifest 
themselves. The scale dependence of monopole contributions to the string
tension was studied in \cite{StackWen}.

\section*{Acknowledgements}
SJH is supported by a
PPARC Advanced Fellowship. The work of JDS and RJW 
was supported in part by the U.S. National Science Foundation 
under grants PHY-9412556 and PHY-9403869. JDS and RJW also thank the
Higher Education
Funding Council for Wales for support in the early stages of this work.
We thank John Mehegan for help in analyzing the fermion propagator.

\begin{table}[pt]
\caption{The values of $\langle\bar\psi\psi\rangle={\pi \over
V}\rho(0)$
obtained from linear fits to $\rho(\lambda)$.}
\label{tab:1}
\bigskip
\begin{tabular}{|c|cl|}
$\beta$  & $\langle\bar\psi\psi\rangle^{U(1)}$ &
$\langle\bar\psi\psi\rangle^{
mon}$\\
\hline
1.005 & 0.165(3)  & 0.105(1)\ \\
1.010 & 0.138(2) & 0.089(2)\  \\
1.020 & 0.002(1)  & 0.0009(6)\\
\end{tabular}

\bigskip

\caption{Results of the fits $m_R (m_0)$ (cf. (\ref{eq:xxx}))}
\label{tab:2}
\bigskip
\begin{tabular}{|c|cc|}
$\beta$  & {$Z$}  & $b$  \\
\hline
1.005 & $1.570(4) $  & $0.0124(3)$ \\
1.010 & $1.561(5) $  & $0.0126(3)$ \\
1.020 & $1.524(4) $  & $0.0115(3)$ \\
\end{tabular}

\bigskip

\caption{Comparison of our two determinations of the renormalization
factor $Z$; The column labeled $Z_m$ comes from the photon
configurations via Eq.~(\ref{eq:xxx}), and the column labeled
$Z_{\bar\psi\psi}$ comes from the chiral condensates via
Eq.~(\ref{eq:psi})}

\label{tab:3}
\bigskip
\begin{tabular}{|c|cc|}
$\beta$  & {$Z_m$}  & $Z_{\bar\psi\psi}$\\
\hline
1.005 & 1.570(4) & 1.570(30)\\
1.010 & 1.561(5) & 1.550(40)\\
\end{tabular}
\end{table}

\begin{figure}[hbt]
\bigskip
\epsfbox{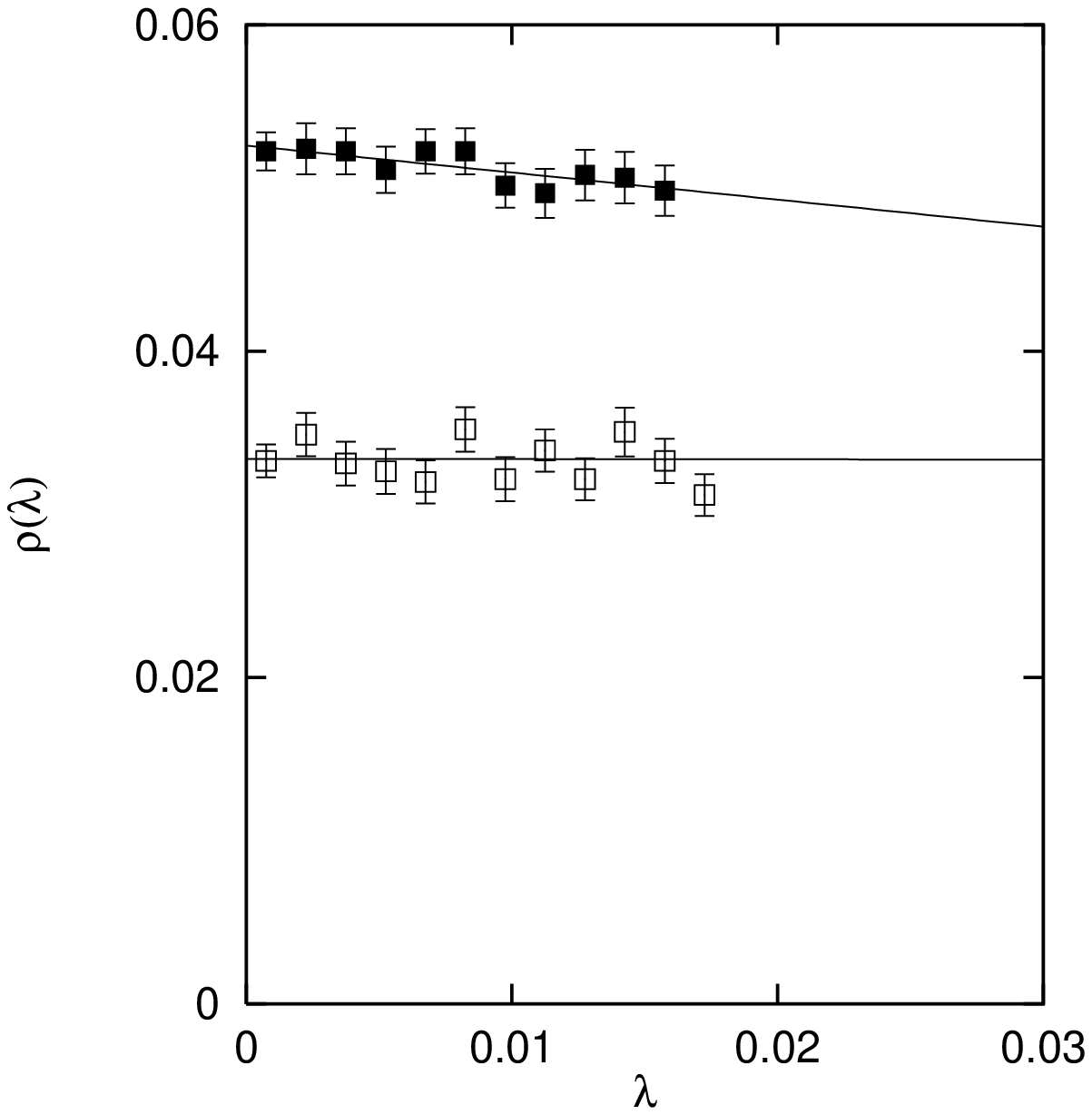}
\caption{The spectral density function for $\beta=1.005$ from
full $U(1)$ fields (solid squares) and monopoles (open squares). }
\end{figure}

\begin{figure}[hbt]
\bigskip
\epsfbox{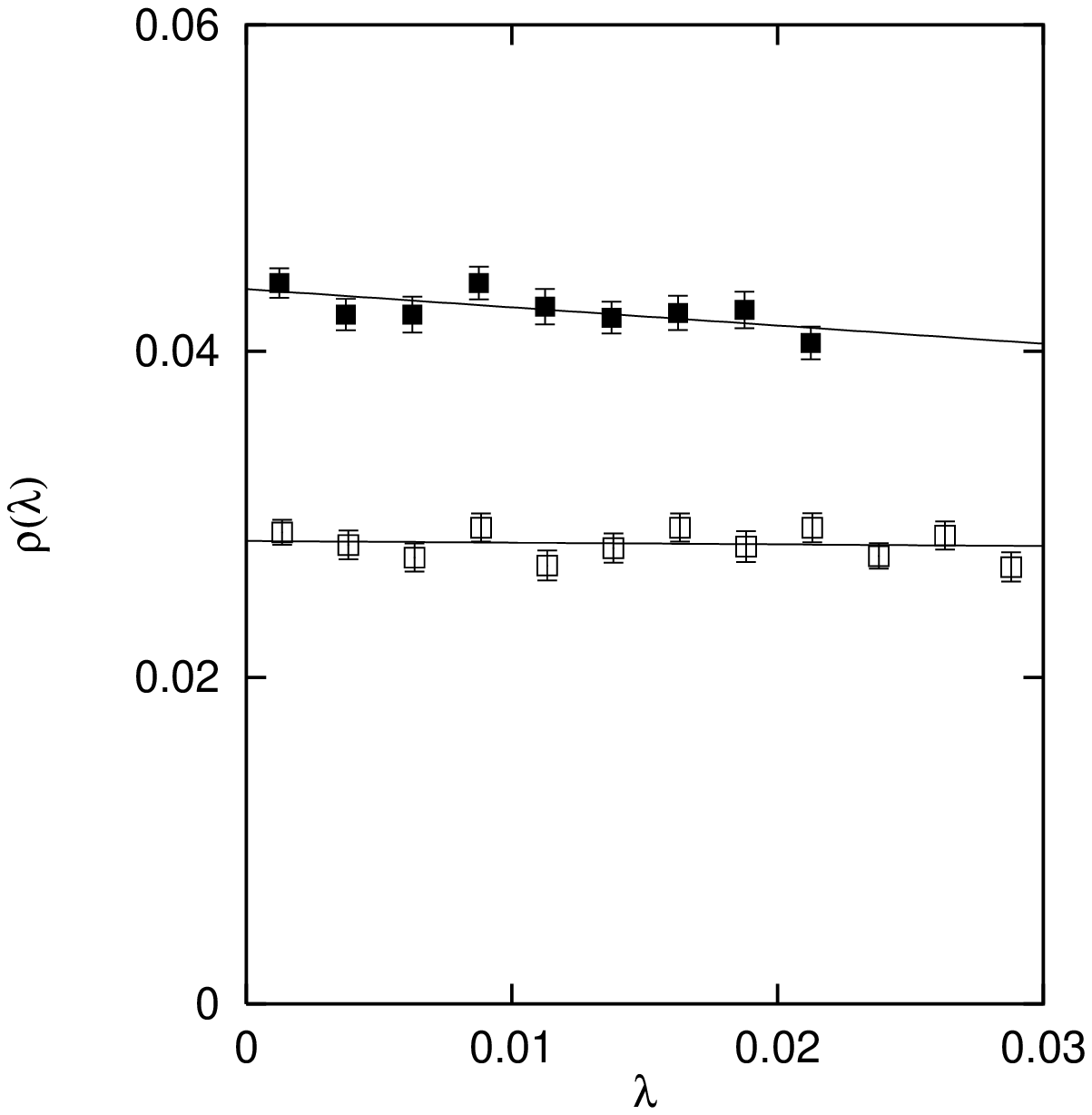}
\caption{The spectral density function for $\beta=1.010$ from
full $U(1)$ fields (solid squares) and monopoles (open squares). }
\end{figure}

\begin{figure}[hbt]
\bigskip
\epsfbox{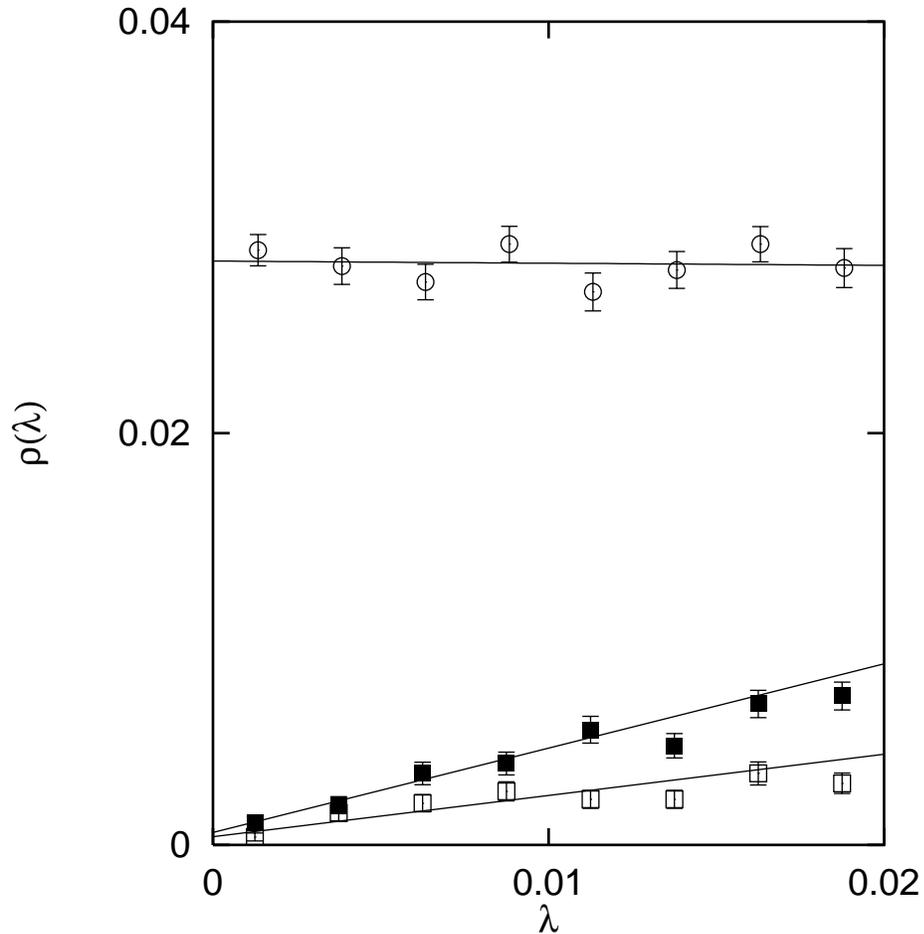}
\caption{The spectral density function for $\beta=1.020$ from
full $U(1)$ fields (solid squares) and monopoles (open squares).
For comparison the results from monopoles for $\beta=1.010$ (open
circles) are
also plotted. }
\end{figure}

\begin{figure}[hbt]
\bigskip
\epsfbox{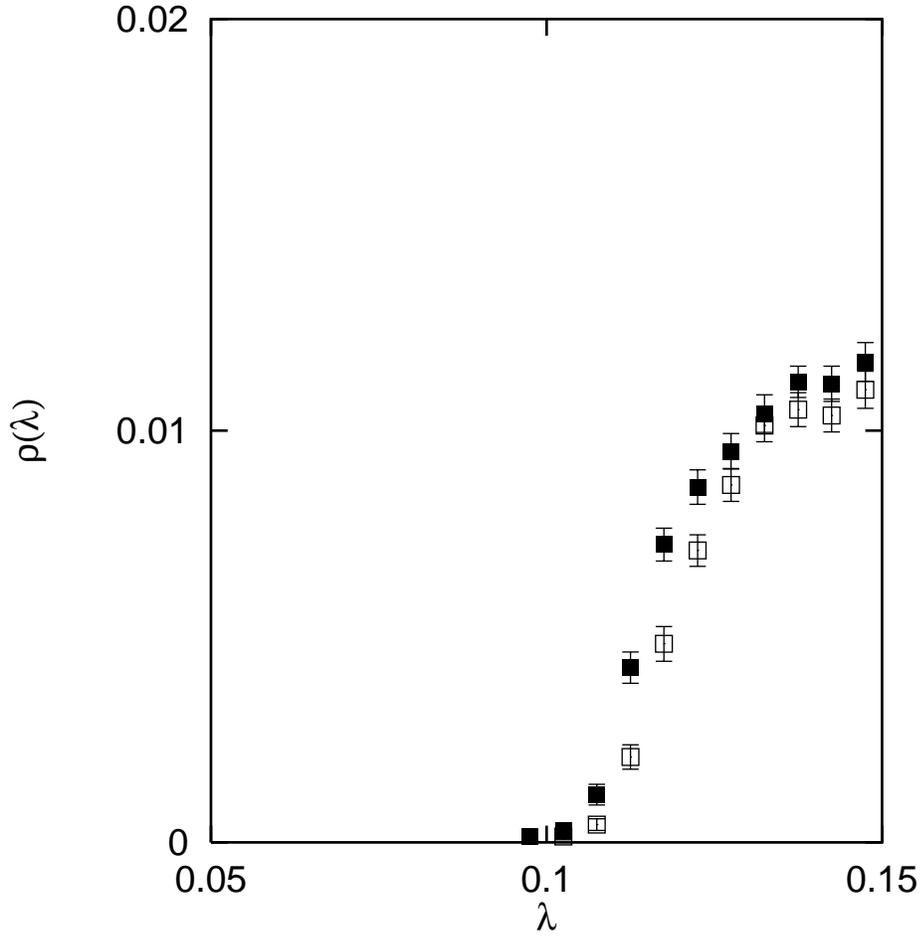}
\caption{The spectral density function from
the photon background $\{U_\mu\}^{phot}$ for $\beta=1.005$ 
(solid squares) and $\beta=1.020$ (open squares).}
\end{figure}

\begin{figure}[hbt]
\bigskip
\epsfbox{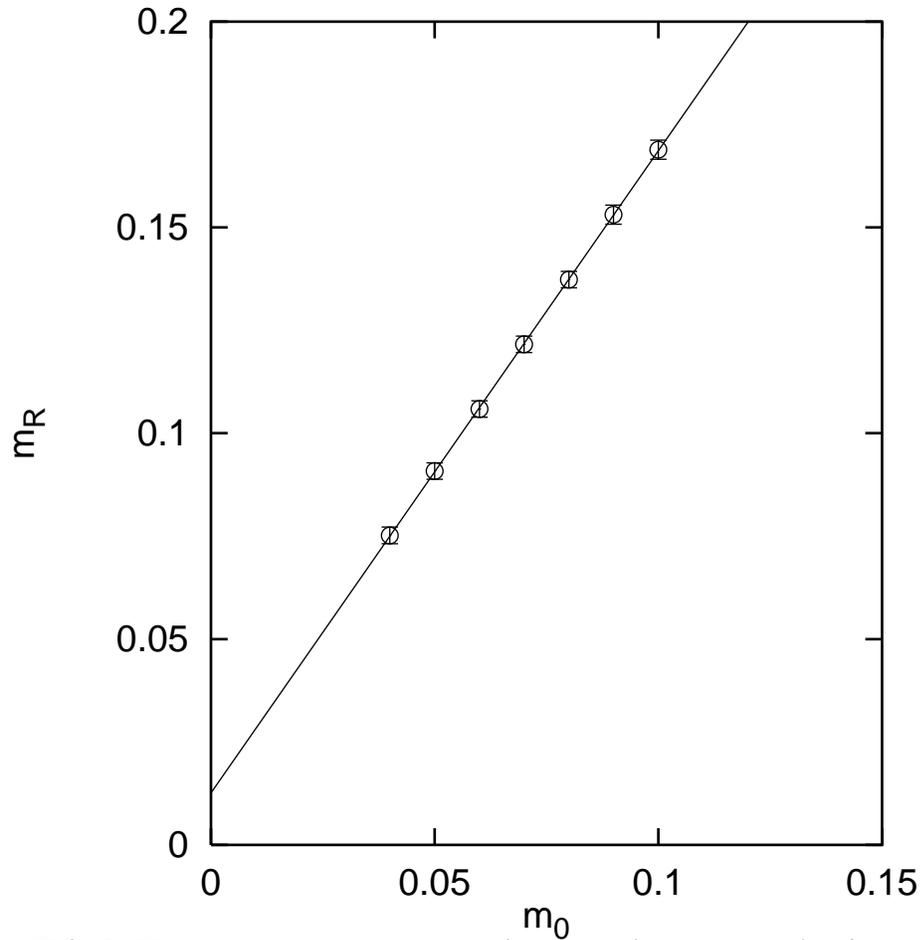}
\caption{The renormalized mass as a function of input mass for
$\beta=1.010$ calculated using the photon propagator
as defined in Eqs.~(14) and~(15).}
\end{figure}

\end{document}